\documentclass[12pt,a4paper]{article}
\usepackage{latexsym}
\usepackage{amsmath}

\begin{document}

\begin{center}  {\Large  \bf
Nonzero cosmological constant\\[2mm]
and the many vacua world}
\end{center}

\vspace{1mm}
\begin{center}
\bf A. A. Grib
\end{center}
    \begin{center}           {\small  \it
A.\,Friedmann Laboratory for Theoretical Physics,     \\
30/32 Griboedov can, St.Petersburg, 191023, Russia  \\
    andrei\_grib@mail.ru  }
\end{center}

\begin{abstract}
    The idea of the quantum state of the Universe described by some
density matrix, i.e mixture of at least two vacua, the trivial symmetric
and the nontrivial one with spontaneously broken symmetry is discussed.
Nonzero cosmological constant necessarily arises for such a state and
has the observable value if one takes the axion mass for the vacuum
expectation value.
The Higgs model, Nambu's model and discrete symmetry breaking are
considered.
Human observers can observe only the world on the nonsymmetric vacuum,
the world on the other vacuum is some dark matter.
Gravity is due to action of two worlds.
Tachyons nonobservable for visible matter can be present in the dark
matter, leading to some effects of nonlocality in the space of
the Universe.
\end{abstract}

{\centering \section{Introduction}}

    Any theory of spontaneously  broken symmetry, the most known being
the Higgs model, leads to the possibility of  at least two
vacua.
One vacuum corresponds to the trivial solution, the other to the
nontrivial one with nonzero vacuum expectation value of the
Higgs field.
If one considers the vacuum expectation value of the quantized stress
energy tensor and puts it into the right hand side of Einstein's
equations one will see that two vacua are different by the nonzero
cosmological constant appearing for the nontrivial
vacuum.
The observable value of the cosmological constant can be obtained by taking
vacuum expectation value of the order of one thousand's of eV,
i.e. the mass of some axion.
So if one puts the hypothesis, expressed by the author many years ago,
that the quantum state of the Universe is not a pure state but the mixture
of two vacua --- the normal and the superfluid one, nonzero cosmological
constant necessarily is observed.
Human observers can observe only the world on the nontrivial vacuum but
gravitation is influenced by worlds on both vacua.
Different vacua correspond to unitary nonequivalent representations of
commutation relations and to orthogonal Hilbert spaces of
states.
So differently from other situations in physics in cosmology if the
volume is infinite  no tunneling is possible from one vacuum to
the other.

    Three models of spontaneous breaking of symmetry are
considered: discrete symmetry breaking with one real scalar
field, the Higgs model for the complex scalar field and the
Nambu model with spinor particles.
In the first model to the massive real field on one vacuum correspond
tachyons on the other.

     For Higgs model one predicts existence of the relativistic dark
matter composed of massless  vector particles and tachyons.
For the Nambu model  to massive fermions and Goldstone massless particles
on one vacuum correspond massless fermions of hidden matter on the other.
Differently from the domains idea these vacua exist everywhere due to
translational invariance and no domain walls are present in the Universe.

    Tachyons are nonobservable in the world where human observers exist,
they are massive Higgs bosons in this world.
     However their existence in the other vacuum world due to gravitational
effects can lead to correlations on spacelike intervals for ordinary matter
which can solve some Friedmann cosmology paradoxes.
The main result of the mixture of vacua hypothesis is that if dark matter
correspond to the world on the other vacuum then the nonzero
cosmological constant must be observed.

\vspace{9mm}
\section{Nonzero cosmological constant in spontaneously broken
symmetry theories}
\hspace{\parindent}
    Observations show that the cosmological constant in the observable
Universe is some positive nonzero number.
    The widespread opinion is that  nonzero cosmological constant
can arise due to special properties of the vacuum for quantum
particles in the Universe.
    What  are these properties and how they can lead to the observable
value of the cosmological constant?
    There are two different ways to obtain this constant
from the properties of the quantum vacuum.

1. If one calculates the vacuum expectation value of the stress-energy
tensor for free quantized field in curved space-time~\cite{GMM}
one obtains some divergent expression.
    The leading divergent term is proportional
to the fourth degree of the momentum and  has the form of the
cosmological term in Einstein's equations.
     This term is present also in flat Minkowski space-time and
describes quantum vacuum oscillations.
    The cosmological constant occurs to be positive
for bosons and negative for fermions.
   This gives hope that due to some supersymmetry for bosons and fermions
the divergence can be cancelled and some finite value for the cosmological
constant can occur.
    However there is no quantitative method to obtain
this value, so the whole reasoning has only some qualitative
meaning.

    2. The other way to obtain the nonzero value of the
cosmological constant as it was shown by the author in 1967~\cite{2}
is to consider the models with spontaneous breaking of discrete
and continuous symmetries in quantum field theory in curved
space-time and consider the quantum state of the Universe to be
not a pure but a mixed state.
     Instead of taking the so called
wave function of the Universe as a pure state one can consider
some mixture of vacua, the trivial and the nontrivial one
described by some density matrix with weights for elements of
the mixture.

    In these models one has the Lagrangian and Hamiltonian invariant
under some symmetry transformations but there exist two different vacua,
the trivial  invariant one and the other noninvariant vacuum.
   For the  continuous symmetry breaking one has the degenerate
noninvariant vacuum and infinite number of different vacua with the same
energy can be obtained by the symmetry transformation.
   In this case Goldstone massless bosons exist which however can acquire
nonzero mass if some explicit breaking of symmetry in the
Lagrangian is taken into account.

    a) Take the simplest model of one real  scalar or pseudoscalar field
with selfinteraction with the Lagrangian
    \begin{equation}
L= - \frac{1}{2} \left( \frac{\partial \Phi}{\partial x_\mu} \right)^2
+ \frac{1}{2}\, \mu_0^2\, \Phi^2 - \frac{\lambda}{4}\, \Phi^4 \,.
\label{1}
\end{equation}
     The wrong sign  for mass squared means that the free field corresponds
to some tachyons, which is the usual case for Goldstone models.
    The potential
\begin{equation}
V(\Phi) = \frac{1}{2}\,\mu_0^2\, \Phi^2 - \frac{\lambda}{4}\, \Phi^4
\label{2}
\end{equation}
    has two  extremal points defined by the equation
\begin{equation}
\mu_0^2\, \Phi - \lambda\, \Phi^3 = 0
\label{3}
\end{equation}
    so that besides the trivial zero solution  the nontrivial solution
also exists.
    In quantum field theory these two solutions correspond to two
different vacua, one with the zero vacuum expectation value, the other
one with the nonzero value   $\Phi_0=\mu_0/\sqrt{\lambda}$.
     For pseudoscalar particles nonzero value means that $P$-invariance
of the Lagrangian is broken for the nontrivial vacuum.
     Quantization on the nontrivial vacuum is made by going to new
fields obtained by the shift on the vacuum expectation value of
the field, so that in terms of the new fields taking into
account the equation for the vacuum expectation value one gets
a new Lagrangian
    \begin{equation}
L = - \frac{1}{2} \left( \partial_\mu \varphi \right)^2 -
\frac{1}{2}\, \mu_0^2\, \varphi^2 - \lambda \Phi_0\, \varphi^3 -
\frac{\lambda}{4}\, \varphi^4 + C
\label{4}
\end{equation}
    where for the constant $C$  one has $ C = \mu_0^4/ (4 \lambda) $.
     If one considers vacuum expectation value of stress-energy
tensor in the normal form of the quantized scalar (pseudoscalar) field
due to translational invariance of vacuum
(in curved space in homogeneous space-time instead of Minkowski
translations one must consider elements of the group of homogeneoity)
this vacuum expectation value is equal with the negative sign to the
vacuum expectation value of the potential.
    So for the nontrivial vacuum one obtains in the right hand side of
Einstein's equations after multiplying on the gravitational constant
some positive cosmological constant $ \Lambda = G\, C $.
    Its numerical value is calculated if one knows the values of
the mass of the boson and the selfinteraction constant.
    For the trivial vacuum this cosmological constant is zero.
    However one can speculate that the Lagrangian and the Hamiltonian are
defined up to some constant, so it may be that this arbitrary
constant is defined so that it just compensates the new constant
arising on the nontrivial vacuum.
    The idea expressed by the author in~\cite{2} was that in cosmology
one can think that both vacua are realized in the infinite Universe with
some weights in the density matrix of the Universe.
    Hilbert spaces in which unitary nonequivalent representations of
commutation relations for the scalar field can be realized are orthogonal.
    So no quantum physical process can connect these states.
    For the actually infinite volume of the Universe
(however it can be that this volume must be much larger than the volume
inside the horizon!) no tunneling from one vacuum to the other is possible.
    This is the difference of the situation in the Universe and that
arising for any finite volume manybody system when tunneling
with nonzero probability is possible.
    That is the reason that no domains or space regions with one and
the other vacua can coexist.
    Both vacua are translational invariant in our case.
    There are two disconnected different worlds.
    No physical interaction exists between them.
    However our hypothesis is that both worlds gravitate and it is only
due to the gravity
(quantization of which is still nonexistent) that they interact.
    One of the worlds is considered from the point of the other as
some dark matter and the nonzero cosmological  constant must be
necessarily present in Einstein's equations!
    If one takes selfinteraction constant for scalar particles in our
model to be of the order of unity, then observational value of the
cosmological constant is consistent with the mass of our field
of the order of $10^{-3}$\,eV, i.e. of the axion in some models~\cite{3}.
    In the result for our simple model  one can come to the
following conclusions.
    We as human observers live on one nontrivial vacuum where no tachyons
can be observed, instead we can observe massive scalar (pseudoscalar)
particles with the axion mass.
    These particles can interact with other particles
which have interactions with the electromagnetic field forming
the observable Universe.
    However there exists the other world on the trivial vacuum,
nonobserved by us through electromagnetic interaction and where we cannot
have our ordinary body made from usual particles.
   So anthropic principle makes this world unseen and we cannot be
conscious of it due to the usual nervous mechanism.
   In this other world tachyons as some dark matter exist.
   Tachyon theorists~\cite{6Recami}  claim that tachyons can account
for some correllations on spacelike intervals.
    Friedmann cosmology for the early Universe has the well known problem
of causality which inflation tries to solve.
    But if tachyons exist on the other vacuum they can provide
another solution of the problem.
    In our reasoning we for simplicity took both weights for vacua of
the order of unity.
    However one can take one of these weights much smaller
than the other, so that the mass of the scalar particle can be
much larger.
    In nonstationary Friedmann Universe these weights
can depend on time which opens the way to new theoretical possibilities.

    b) The Higgs model.
    One has the complex scalar field (or equivalently two real fields)
with tachyonic mass (negative sign of the mass squared) interacting with
massless vector field and with selfinteraction term as in the previous
case with the Lagrangian
    \begin{equation}
L = - \frac{1}{2} \left( \nabla \varphi_1 \right)^2 -
\frac{1}{2} \left( \nabla \varphi_2 \right)^2 -
V ( \varphi_1^2 + \varphi_2^2 ) -
\frac{1}{4}\, F_{\mu \nu} F^{\mu \nu}
\label{5}
\end{equation}
       where
    \begin{equation}
\nabla_\mu \varphi_1 = \partial_\mu \varphi_1 - e_0 A_\mu\, \varphi_2 \,,
\ \ \ \ \ \
\nabla_\mu \varphi_2 = \partial_\mu \varphi_2 + e_0 A_\mu\, \varphi_1 \,,
\label{6}
\end{equation}
    \begin{equation}
F_{\mu \nu} = \partial_\mu A_\nu - \partial_\nu A_\mu \,.
\label{7}
\end{equation}
    Looking for the stationary points of the potential
    \begin{equation}
V ( \varphi_1^2 + \varphi_2^2 ) =
\mu_0^2 \left( \varphi_1^2 + \varphi_2^2 \right)
- \frac{1}{6}\, \lambda \left( \varphi_1^2 + \varphi_2^2 \right)^2
\label{8}
\end{equation}
    one obtains two solutions
    \begin{equation}
\varphi_1 = \varphi_2 = 0 \,; \ \ \ \ \ \ \
\varphi_1 = 0 \,, \ \ \    \varphi_2 = \varphi_0 \ne 0 \,.
\label{9}
\end{equation}
     The solution leading to noninvariant nonzero vacuum expectation
value describes one scalar field of the Higgs boson with the mass,
the other scalar component forms the degree of freedom
of the vector massive boson so that one comes to equations
    \begin{equation}
\left( \Box - 4 \varphi_0^2\, V''(\varphi_0^2) \right)
{\tilde{\varphi}}_2 =0 \,,
\label{10}
\end{equation}
    \begin{equation}
\partial_\nu F^{\mu \nu} - e_0^2\, \varphi_0^2\, B^\mu =0 \,,
\label{11}
\end{equation}
    \begin{equation}
B_\mu = A_\mu - ( e_0 \varphi_0)^{-1} \partial_\mu {\tilde{\varphi}}_1
\label{12}
\end{equation}
      As in the model discussed by us previously the nonzero
cosmological constant is equal to
    \begin{equation}
\Lambda = G\, \frac{1}{6} \lambda\, \varphi_0^4 \,.
\label{13}
\end{equation}

    c) Nambu's model for the spinor field.
    The Lagrangian is
    \begin{equation}
L = \overline{\psi}\, i\, \partial_\mu \nabla^\mu \psi + g \left[ (
\overline{\psi} \psi )^2 - ( \overline{\psi} \gamma_5 \psi)^2 \right].
\label{14}
\end{equation}
    There are two different vacua for this chiral invariant Lagrangian.
Following~\cite{4}   define
    \begin{equation}
m_S(x) = 2\, g\, i \, {\rm Tr}\, S_F(x,x)\ , \ \ \ \ \
m_P(x) = 2\, g \, {\rm Tr}\, ( \gamma_5 S_F(x,x)) \,,
\label{15}
\end{equation}
    \begin{equation}
\Phi(x) = m_S(x) + i m_P(x).
\label{16}
\end{equation}
     Then one obtains the analog of the Ginzburg-Landau equation
for the superconductor wave function
    \begin{equation}
\Box \, \Phi - 2 m^2_\infty \Phi (x) + 2 | \Phi(x)|^2 \Phi(x) =0
\label{17}
\end{equation}
    where
\vspace{-9mm}
    \begin{equation}
\Phi(x) \xrightarrow[x \to \infty]{}   m_\infty \,.
\label{18}
\end{equation}

    As it is well known in Minkowski space-time one must introduce
some cutoff parameter for the theory to be consistent.
    However in curved space-time with the negative curvature of space one
can have nontrivial vacuum without the cutoff~\cite{5}.
    Two vacua, the trivial and the nontrivial  one are different  by the
cosmological constant
            \begin{equation}
\Lambda = G \left\langle 0'\! \left|\, g \left[ ( \overline{\psi} \psi )^2
- ( \overline{\psi} \gamma_5 \psi)^2 \right] \right| 0' \right\rangle.
\label{19}
\end{equation}
      This constant can have different sign depending on the values of
different vacuum expectation values $\ m_S, \ m_P$.\
    This means that there is the possibility of compensation of
different cosmological constants occurring  due to Higgs bosons
etc and fermions in the Nambu model.
    Two worlds are different in the sense that massless fermions with
exact chiral symmetry exist in one world, while in the other world
there exist massive fermions and Goldstone pseudoscalar particles.
    If one takes the observable value of the cosmological constant then
for close to unity weights in the density matrix for vacua the
mass of the fermion must be close to the axion mass discussed
before, so that one can identify it as some neutrino particle.
    Can we make some estimates on weights in the density matrix for vacua?
    If  the world on the trivial vacuum is identified with some dark matter
then it seems that in the present form our model must lead to the
identification of cold dark matter with massive tachyons and to
the difference in weights consistent with observations not larger
than one percent.
    This excludes the possibility of taking instead of the axion mass
the Higgs boson mass of the electroweak model.
    For Higgs model the weight of the nontrivial vacuum at the
modern epoch must be very small.

\vspace{4mm}
{\bf Acknowledgements}.
    This work was supported by Min. of Education of Russia,
grant E02-3.1-198.



\begin{thebibliography}{4}

\bibitem{GMM}
A.\,A. Grib, S.\,G. Mamayev and V.\,M. Mostepanenko,
{\it Vacuum Quantum Effects in Strong Fields.}
Energoatomizdat, Moscow, 1988.
[English transl.: Friedmann Lab. Publishing, St.Petersburg, 1994.]

\bibitem{2}
A.\,A. Grib,
{\it $CP$-noninvariance in $K^0$-meson decays and
     nonequivalent representations in quantum field theory.}
Vestnik Leningrad. University {\bf 22}, (1967) 50--56.

\bibitem{3}
A.\,D. Chernin,
{\it Cosmic vacuum.}
Uspekhi Fiz. Nauk {\bf 171}, (2001) 1153--1175.
[English transl.: Physics -- Uspekhi {\bf 44}, (2001) 1099--1118.]

\bibitem{6Recami}
E. Recami,
{\it Relativity and beyond.}
{In } ``Albert Einstein, 1879--1979, Relativity, Cosmology and Quanta",
Vol.~2, New York, 1979, p.~537--597.

\bibitem{4}
T. Eguchi, H. Sugawara,
{\it Extended model of elementary particles based on an analogy with
     superconductivity.}
Phys. Rev.~D {\bf 10}, (1974) 4257--4262.

\bibitem{5}
A.\,A. Grib,
{\it Higgs's phenomenon without Higgs's mesons.}
Izvestiya Vuzov, Physics, N~{\bf 9}, (1981) 129--130.

\end{thebibliography}
\end{document}